\documentclass[prl,twocolumn,amsmath,amssymb,superscriptaddress,showpacs]{revtex4}
\usepackage[latin9]{inputenc}
\usepackage{amsmath}
\usepackage{amssymb}
\usepackage{epsfig}

\usepackage{ucs}
\usepackage{bbm}

\newcommand{\bra}[1]{\langle #1|}
\newcommand{\ket}[1]{|#1\rangle}
\newcommand{\ketbra}[1]{| #1\rangle \langle #1|}

\newcommand{\be}{\begin{equation}}
\newcommand{\ee}{\end{equation}}
\newcommand{\eea}{\end{eqnarray}}
\newcommand{\bea}{\begin{eqnarray}}

\newcommand{\va}[1]{\ensuremath{(\Delta#1)^2}}

\newcommand{\exs}[1]{\ensuremath{\langle{#1}\rangle}}
\newcommand{\mean}[1]{\ensuremath{\langle{#1}\rangle}}

\newcommand{\qed}{\ensuremath{\hfill \blacksquare}}

\newcommand{\kommentar}[1]{}
\newcommand{\trace}{{\rm Tr}}

\newcommand{\forget}[1]{}
\newcommand{\tr}{\mbox{Tr}}

\newcommand{\elinprime}{E_{\rm lin}^{\rm (ppt)}}

\newcommand{\Coperator}{M}
\newcommand{\ckw}{\tau}

\newcommand{\omegassep}{\omega_{12} \text{ symmetric, separable}}

\newcommand{\omegasppt}{\omega_{12} \text{ symmetric, PPT}}

\newcommand{\comment}[1]{}
\newcommand{\ifrac}[2]{{#1}/{#2}}

\begin{document}

\title{Evaluating convex roof entanglement measures}

\author{G\'eza T\'oth}
\email{toth@alumni.nd.edu}
\homepage{http://www.gtoth.eu}
\affiliation{Department of Theoretical Physics, University of the Basque Country
UPV/EHU, P.O. Box 644, E-48080 Bilbao, Spain}
\affiliation{IKERBASQUE, Basque Foundation for Science, E-48011 Bilbao, Spain}
\affiliation{Wigner Research Centre for Physics, Hungarian Academy of Sciences,
P.O. Box 49, H-1525 Budapest, Hungary}

\author{Tobias Moroder}
\affiliation{Naturwissenschaftlich-Technische Fakult\"at, Universit\"at Siegen, Walter-Flex-Str. 3, 57068 Siegen, Germany}
\author{Otfried G\"uhne}
\affiliation{Naturwissenschaftlich-Technische Fakult\"at, Universit\"at Siegen, Walter-Flex-Str. 3, 57068 Siegen, Germany}


\begin{abstract}
We show a powerful method to compute entanglement measures based
on convex roof constructions. In particular, our method is applicable
to measures that, for pure states, can be written as low order polynomials of
operator expectation values. 
We show how to compute the linear entropy of entanglement, the linear entanglement of assistance, 
and a bound on the dimension of the entanglement for bipartite systems.
We discuss how to obtain the convex roof of the three-tangle for three-qubit states.
We also show how to calculate the  linear entropy of entanglement and the quantum Fisher information based on
partial information or device independent information.
We demonstrate the usefulness of our method by concrete examples.
\end{abstract}

\date{\today}

\maketitle

Quantum entanglement plays a central role in quantum information science and quantum optics \cite{GT09}. There are now efficient methods to detect entanglement, that have even been used in many experiments \cite{expdet}. These mostly answer the yes or no question: "Is the quantum state entangled?" or "Is the quantum state genuine multipartite entangled?" After verifying the presence of entanglement, 
the next step is quantifying it.
Calculating measures is 
becoming increasingly important in experiments in quantum information science \cite{GR08, EB07,WP09} and it also plays a crucial role in investigations in quantum statistical physics, e.g., in studying phase transitions \cite{OA02}.

Most entanglement measures
 are based on the convex roof of a quantity on pure states such as the entropy of 
 the reduced state \cite{WK02,WG03,CK00}.
Measures of this type can also be used to classify states according to their membership in some convex sets, 
 for example, based on their Schmidt rank \cite{AB01,SB01}. They 
play a central role in quantum information theory, however, in most of the cases they are not computable as there are no efficient ways to calculate convex roofs. 
Most importantly, the simplest multipartite entanglement measure, the three-tangle for three-qubits, cannot be computed for a general state.

Thus, for obtaining entanglement measures in theory and experiments,
it would be crucial to find methods to calculate
convex roof constructions efficiently, at least for not too large systems.
This seems to be a  very difficult task since straightforward numerical search 
means an optimization over an infinite number of convex
decompositions of the density matrix. Such an approach 
will lead to an
{\it upper} bound on the measure, since a multivariable numerical optimization is not guaranteed to find the global optimum \cite{RL09}. 
Upper bounds, however, are often not very useful as the amount of entanglement can be much lower or even zero even if the procedure signals considerable entanglement. 

In this paper, we present a method that produces a series of very good {\it lower} bounds on important entanglement measures. Our method has the following characteristics: (i) 
It is based on semidefinite programming.
The series of bounds obtained converge in a controllable way to the true value.
Even the first lower bound in the series is non-trivial.
 (ii) We have a clear physical picture for what states our method yields a nonzero value for the measures.
 (iii) The set of separable states is used in the optimization procedure.
 This way we connect calculating convex roofs to the separability problem,
 which might help to find applications of the separability problem in other areas of physics. 
We will demonstrate the use  of our method 
with the example of computing bipartite entanglement measures for bound 
entangled states, computing the convex roof of the tangle for various three-qubit states, and even quantities
outside of quantum information science.
Our method can also be used to compute a
lower bound from incomplete data of the quantum state or in device independent scenarios
 \cite{VMMW02,LV11,MB14,pusey}.

\emph{Convex roof of linear entropy}. 
For pure states, the linear entropy of entanglement  is given as
\be\label{Ex1}
E_{\rm lin}(\ket{\Psi})=S_{{\rm lin}}[{\rm Tr}_1(\ket{\Psi})],
\ee
where we used the definition of the linear entropy 
$S_{\rm lin}(\varrho)=1-{\rm Tr}(\varrho^{2}).$ 
Hence, the linear entropy of entanglement for pure states 
equals also $C^2/2,$ where $C$ is the concurrence \cite{WK02},
and it is also equal to the $I$-tangle \cite{Itangle}.
The definition \eqref{Ex1} can be extended to mixed states
by a convex roof construction 
as
\begin{equation} \label{Elininf1}
E_{\rm lin}(\varrho)=
 \min_{\{p_{k},\ket{\Psi_{k}}\}}\bigg(
 \sum_{k}p_{k}
E_{{\rm lin}}(\ket{\Psi_{k}})\bigg),
\end{equation}
where $\{p_{k},\ket{\Psi_{k}}\}$ is a decompositon to pure states 
\be\varrho=\sum_{k}p_{k}\ketbra{\Psi_{k}}.\label{Elininf}\ee
It can be shown that  $E_{\rm lin}(\varrho)$  does not increase under
local operations and classical communication (LOCC) on average, hence it is an 
entanglement monotone  \cite{elinmonotone}.
Consequently, $E_{\rm lin}(\varrho)$ has also been used to characterize entanglement even in the 
multipartite setting \cite{ElinEnt}.

Next, we will show a method to compute Eq.~(\ref{Elininf1}).
For this aim, first we write the liner entropy of entanglement as an expectation value of an operator acting on two copies of a bipartite pure state 
$\ket{\Psi}$ as \cite{H03}
\begin{equation}\label{Slin}
E_{\rm lin}(\ket{\Psi})=
{\rm Tr}[\mathcal{A}_{\rm AA'} \otimes \openone_{BB'} (\ket{\Psi}\bra{\Psi})_{AB} \otimes (\ket{\Psi}\bra{\Psi})_{A'B'}].
\end{equation}
 Here, $A$ and $B$ denote the parties of the first copy
 while $A'$  and $B'$ denote the parties of the second copy. 
Moreover, the projector to the antisymmetric space is defined as $\mathcal{A}_{\rm AA'}:=(\openone-\mathcal{F})_{\rm AA'},$ $\mathcal{F}$ is the flip operator, 
and we explicitly wrote out $\openone_{BB'}$ for clarity \cite{MK05}.

Next, we will consider mixed states.
Let us assume that $\{\tilde{p}_{k},\ket{\tilde{\Psi}_{k}}\}$ is the decomposition attaining the convex roof. Then, 
for a state with such a decomposition we obtain
\begin{eqnarray}
E_{\rm lin}(\varrho)
&=&\sum_{k}\tilde{p}_{k} E_{{\rm lin}}(\ket{\tilde{\Psi}_{k}})\nonumber\\
&=&\sum_{k}\tilde{p}_{k} \trace(\mathcal{A}_{AA'}\ketbra{\tilde{\Psi}_k}^{\otimes 2})\nonumber\\
&=&\trace(\mathcal{A}_{AA'}\omega_{12}),
\end{eqnarray}
where the state on the two-copy space is defined as
\begin{eqnarray}\label{omega12}
\omega_{12}=\sum_{k}\tilde{p}_{k} \ketbra{\tilde{\Psi}_k}\otimes  \ketbra{\tilde{\Psi}_k}. 
\end{eqnarray}
The density matrix $\omega_{12}$ has three important properties. It is a mixture of product states, i.e., a separable state \cite{W89}.
Moreover, all the pure product components are symmetric. Thus, $\omega_{12}$ is supported on the symmetric subspace. In fact, any symmetric separable states can be written in the form (\ref{omega12}) \cite{SymSep}.
Finally, ${\rm Tr}_2(\omega_{12})=\varrho.$

Hence, we arrive at our first main result. 

\textbf{Observation 1.}---The convex roof of the linear entropy can be written as 
\begin{eqnarray}\label{Elinrho}
E_{\rm lin}(\varrho)=
\min_{
\omega_{12}}&&
\trace(\mathcal{A}_{AA'} \omega_{12} ),\\
\textrm{s.t.}&& 
\omegassep, \nonumber \\
&&  \omega_{1} = \varrho,\nonumber
\end{eqnarray}
where $ \omega_{1}\equiv {\rm Tr_2}(\omega_{12} ).$ 

Observation 1 connects the separability problem of symmetric bipartite states, i.e., answering the question "Is the state entangled?" mentioned in the introduction, to entanglement quantification. In principle, to obtain a lower bound on $E_{\rm lin}(\varrho),$ any necessary condition for separability could be used. We will consider 
the method based on the positivity of partial transpose (PPT) 
\cite{{PPT}} and obtain a lower bound as
\begin{eqnarray}\label{Elinppt}
\elinprime(\varrho)=
\min_{ \omega_{12} }&&
\trace(\mathcal{A}_{AA'} \omega_{12} ),\\
\textrm{s.t.}&&
\omegasppt, \nonumber\\
&& \omega_{1} = \varrho,\nonumber
\end{eqnarray}
Next, we will demonstrate that our method can be used to quantify the entanglement of 
states not detected by the PPT condition, called bound entangled states \cite{bound,H97,boundinteresting}.

{\it Horodecki state.}---We test our method to calculate entanglement measures for the one-parameter family of the $3\times3$ bound entangled state $\varrho_a^{\rm PH}$ introduced by P. Horodecki \cite{H97}.
We mix the state with white noise according to $\varrho_a(p)=p\varrho_a^{\rm PH}+(1-p)\ifrac{\openone}{9}$ and calculate the entanglement as a function of $a$ and $p.$ 
The results can be seen in Fig.~\ref{elin_horodecki}. 
The critical noise for which $\elinprime(\varrho)=0$ agrees with the calculations of Ref.~\cite{MK05} and Ref.~\cite{CM12}. 
We note that we made the computer program calculating $\elinprime(\varrho),$ with all other programs used for this publication,
publicly available \cite{mathworks}.
Other methods for calculating entanglement measures are in Refs.~\cite{CA05,G03}.

\begin{figure}
\vskip-1cm
\centerline{ \epsfxsize4in \epsffile{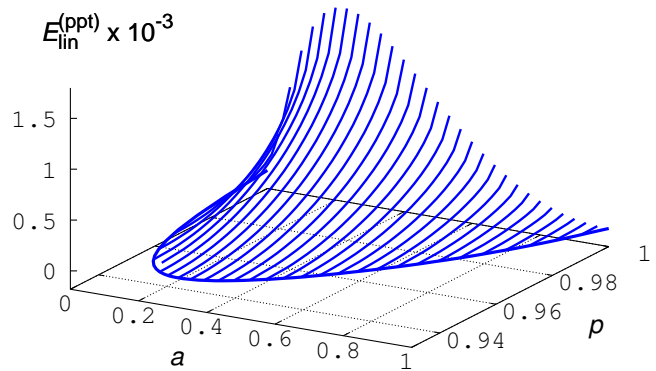}}\vskip-0.7cm
\caption{(Color online)   Entanglement quantification for the noisy $3\times3$ Horodecki bound entangled state $\varrho_a(p)$ using 
$\elinprime(\varrho)$ defined after Observation 1. We plot $\elinprime(\varrho)$ as a function of the parameter of the state, $a,$ and the weight-parameter $p.$ 
}
\label{elin_horodecki}
\end{figure}

It is a surprise that, while the bound relies on the PPT criterion, the method is still able to detect PPT entangled states. 
In order to obtain more information on what kind of states are detected, we need to know the separability criterion based on
symmetric extensions \cite{PPTsymext}. A given bipartite state $\varrho_{AB}$ is said to have a $n:m$ symmetric extension if it can be written as the reduced state of a multipartite state 
$\varrho_{A_1..A_nB_1..B_m},$ which is symmetric under $A_k\leftrightarrow A_l$ and $B_k\leftrightarrow B_l$ for all $k\ne l.$
If we also require that the state is PPT for all bipartitions, then it is a PPT symmetric extension.
Separable states have such extensions for arbitrarily large $n$ and $m,$ while the lack of such  an extension signals the presence of entanglement.

\textbf{Observation 2.}---For all non-PPT states and for all states that do not have a $2:2$ symmetric extension we have 
$\elinprime(\varrho)>0.$
Moreover, for all states having a $2:2$ PPT symmetric extension $\elinprime(\varrho)=0$ holds.
The proof can be found in the Supplement~\cite{SUPP}.

\begin{figure}
 \vskip-0.5cm
\centerline{ \epsfxsize3.8in  \epsffile{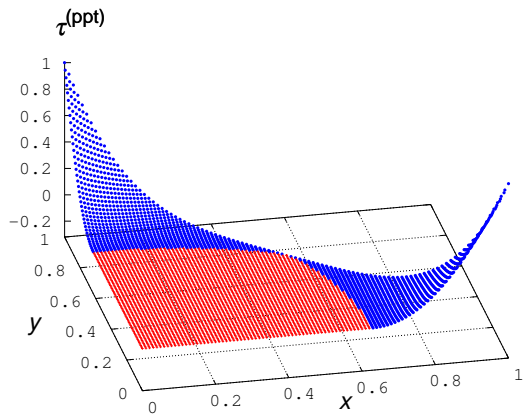}}
\vskip-0.5cm
 \vskip-0.1cm
\caption{(Color online) Three-tangle of a family of states \eqref{rhoxy} as a function of the parameters $x$ and $y.$ Light color indicates the region where the tangle is zero, darker color indicates a nonzero value. 
 }
\label{fig:3tangle}
\end{figure}

Before we continue let us point out that we can also obtain a lower bound on $E_{\rm lin}(\varrho)$ if we choose any other entanglement condition,
such as the method based on local uncertainty relations \cite{lur}, 
the covariance matrix criterion \cite{GH07}, 
or the computable cross norm or realignment
criterion (CCNR) \cite{ccnr}. 
However,  for symmetric states these are all equivalent to the PPT condition \cite{TG09}. 

Therefore, to strengthen the bound a stronger criterion must be employed. 
Here again the method of PPT symmetric extensions can be used \cite{PPTsymext}. 
Rather than approximating $\omega_{12}$ by PPT states, we demand that $\omega_{12}$ has an $n:1$ PPT symmetric extension \cite{symext}.
In this way we obtain a sequence of lower bounds $E_{\rm lin}^{(n)}$ with increasing accuracies. The corresponding optimization can similarly be carried out by semidefinite programming. 
Note that the PPT symmetric extensions converge to the set of separable states in a controlled way \cite{NO09}.
Finally, note also that semidefinite programs not only detect entanglement, but through solving the dual problem, it is possible to find entanglement witnesses \cite{PPTsymext}.
In our case, these witnesses can even bound entanglement measures, as explained in the Supplement  \cite{SUPP}.

{\it  Generalization and further examples.}---The previous ideas can straightforwardly be generalized to compute the convex roof of any quantity that can be written as a polynomial of expectation values for pure states as
\begin{equation}\label{E}
E(\ket{\Psi})=
 \sum_{m=1}^M\sum_{n=0}^Nc_{mn}\exs{A_m}^n,
\end{equation}
where $ A_m$  are operators and $ c_{mn}$  are constants (see e.g., \cite{OS12,G05}). 
It is possible to define an operator $L:=\sum_{m,n}c_{mn}A_m^{\otimes N}\otimes \openone^{\otimes (N-n)},$
whose expectation value on several copies reproduces Eq.~\eqref{E}.
 Then, the convex roof of Eq.~\eqref{E} can be obtained as an optimization over $N$-copy symmetric fully separable states \cite{SymSep}
\begin{eqnarray}
E(\varrho)=
\min_{ \omega_{12..N} }&&
\trace(L \omega_{12..N} ),\\
\textrm{s.t.}&& 
 \omega_{12..N} \text{ symmetric, fully separable}, \nonumber\\
&&\omega_{1} = \varrho.\nonumber
\end{eqnarray}

{\it Three-tangle.}---Our next example is the calculation of the 
three-tangle, a three-qubit entanglement monotone \cite{DV00}. For pure states, it has been defined by Coffmann, Kundu and Wootters \cite{CK00}.
Remarkably, it can be written as a fourth-order polynomial in expectation values \cite{OS12}. 
Hence, for mixed states, the tangle can be defined through a convex roof extension, which we can now map to the optimization problem 
\begin{eqnarray}
\tau(\varrho)=
\min_{ \omega_{12} }&&
\trace(T \omega_{1234} ),\\
\textrm{s.t.}&& 
 \omega_{1234} \text{ symmetric, fully separable}, \nonumber\\
&&\omega_{1} = \varrho,\nonumber
\end{eqnarray}
where $T$ is an operator acting on four copies of the three-qubit state \cite{T}.
Note that if we know $\ckw(\varrho),$ we can decide whether a three-qubit fully entangled state is in the W or in the GHZ class \cite{AB01}.

The optimization can be carried out for symmetric multiqubit states that are PPT with respect to all bipartions rather than 
symmetric separable states, leading to the lower bound $\ckw^{({\rm ppt)}}.$
 The results are shown in Fig.~\ref{fig:3tangle} for states of the form
\bea
\varrho(x,y)&=&x\ketbra{{\rm GHZ}^+}+y\ketbra{{\rm GHZ}^-}\nonumber\\
&&+(1-x-y)\ketbra{{\rm W}},\label{rhoxy}
\eea
where $\ket{{\rm GHZ}^\pm}=(\ket{000}\pm\ket{111})/\sqrt{2},$ and
$\ket{{\rm W}}=(\ket{100}+\ket{010}+\ket{001})/\sqrt{3}.$
Note that a lower bound for the convex roof of 
the tangle for general states, which is exact for states with certain symmetries, has been developed \cite{ES12}.

As a practical comment, we add that the numerical computation is challenging,
but $\ckw^{({\rm ppt)}}$ can be computed on a standard laptop 
with standard free packages for semidefinite programming \cite{SDP}, if the state has some symmetries, or has a rank up to six. Calculations for general three-qubit states of rank eight are realistic with 
computer clusters and professional packages.

\begin{figure*}
 \centerline{
   \epsfxsize 3.2in \epsffile{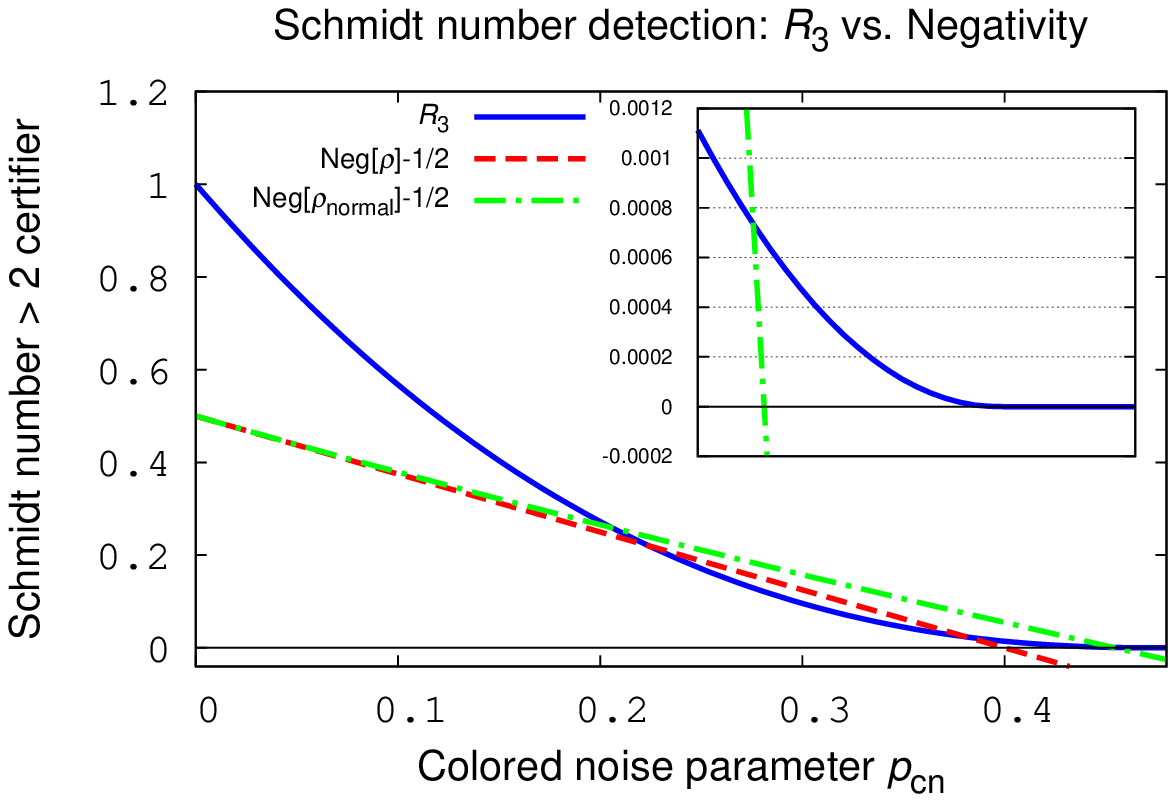}
    \epsfxsize3.2in \epsffile{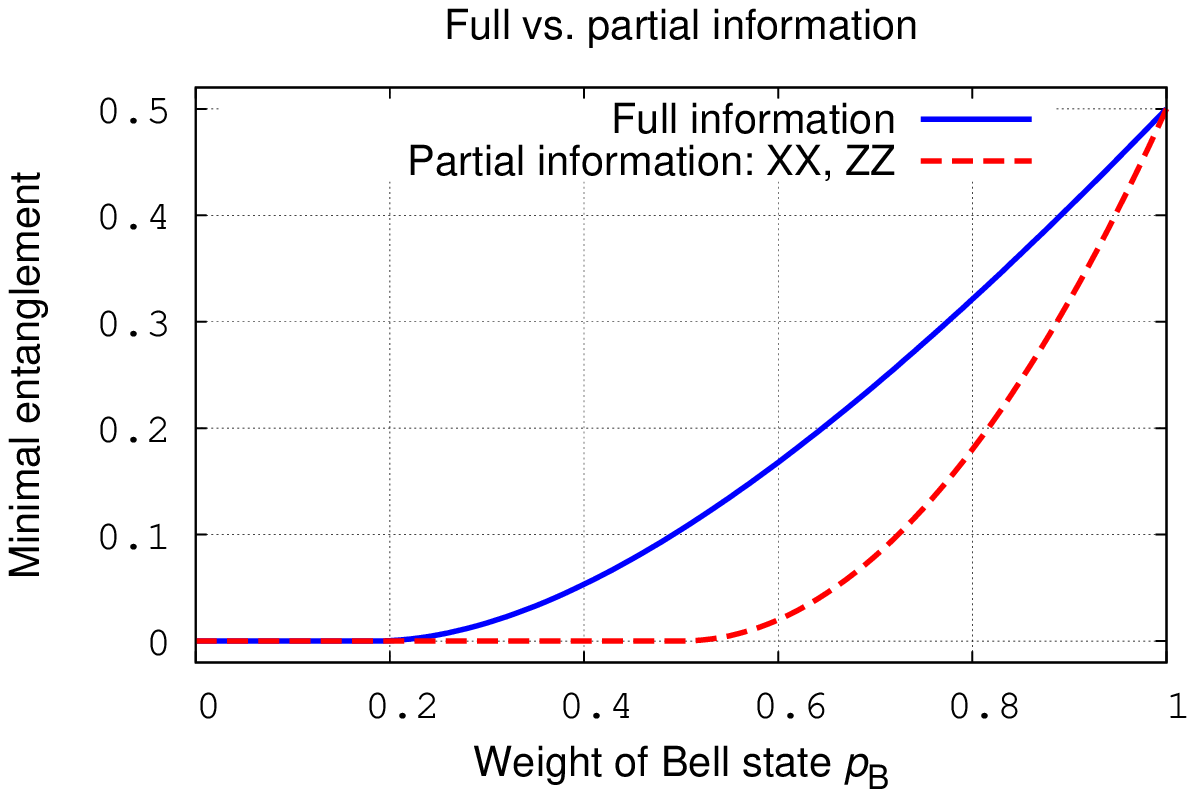}}
 \centerline{(a) \hskip8cm (b)}
 \centerline{   
  \epsfxsize3.2in \epsffile{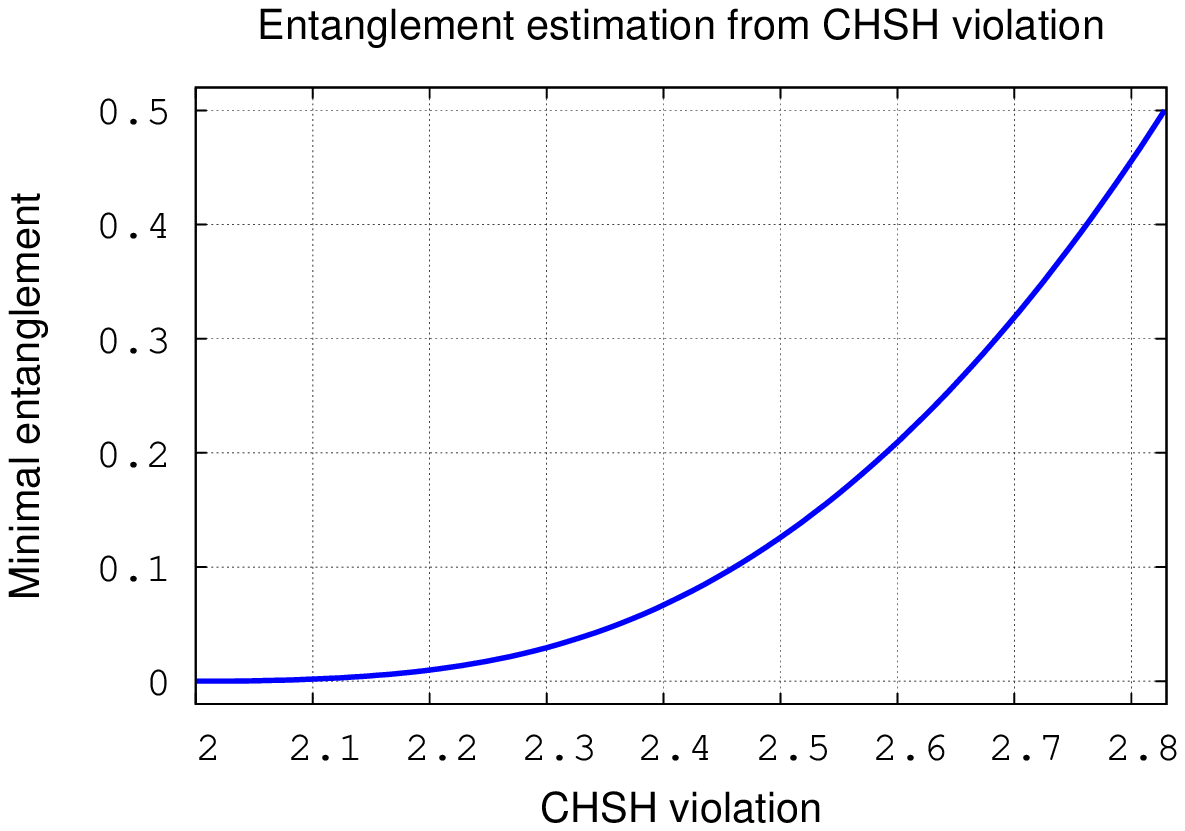}
  \epsfxsize3.2in  \epsffile{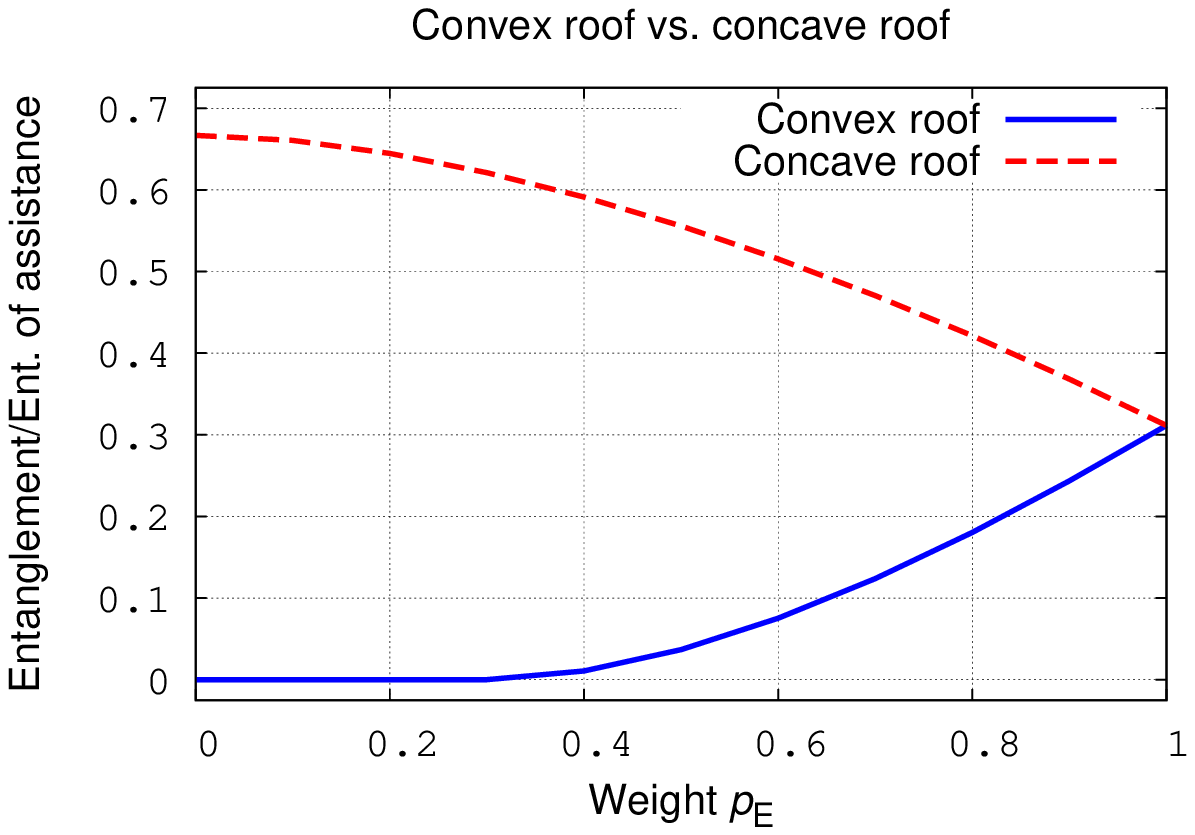}
 }
  \centerline{(c) \hskip8cm (d)}
  \caption{(Color online)  
(a) Schmidt-number witness vs. negativity for a state of the type
\eqref{state_for_Schmidt} as a function of $p_{\rm cn}.$ 
 As the inset shows, even when we consider negativity of the normal form, obtained through stochastic local operations and classical communications (SLOCC) such that all local matrices are fully mixed \cite{normalform}, our numerical method is superior.
(b) $\elinprime(\varrho),$ given in Eq.~\eqref{Elinppt}, based on partial information for the state \eqref{partial}.
(c) Estimation of $\elinprime(\varrho)$ as a function of the violation of the CHSH inequality.
(d) $\elinprime(\varrho)$  and the corresponding bound for the entanglement of assistance (defined with the linear entropy) for a state of the type
\eqref{state_for_Elin} as a function of $p.$ 
}
\label{elin_results}
\end{figure*}

{\it  Schmidt rank.}---Let us consider the quantities $R_r$ that are nonzero for states with a Schmidt rank larger than $r.$ For example,
$R_2(\vert\Psi\rangle)=\sum_{i<j}\lambda_i\lambda_j\equiv 4 S_{\rm lin}(\vert\Psi\rangle),$
$R_3(\vert\Psi\rangle)=\sum_{i<j<k}\lambda_i\lambda_j\lambda_k,$
where $\lambda_k$ are the eigenvalues of the reduced state.
The $R_k$ quantities are proven to be entanglement monotones \cite{SZ02}.
We can calculate the convex roof of $R_r$ with our method.
Convex roofs for such quantities allow us to bound the dimensionality of the entanglement from below. 
A powerful bound can be obtained by carrying out the optimisation
for $r$-qudit symmetric states that are PPT with respect to all bipartitions. 
An alternative is computing the negativity \cite{VW02,ES13}. In particular, $\mathcal{N}(\varrho)-\ifrac{1}{2}>0$ signals that the Schmidt number is larger than $2.$
We show that our method outperforms the negativity as a dimension witness in Fig.~\ref{elin_results}(a)
for the family of states
\bea\label{state_for_Schmidt}
\varrho_{\rm S}(p_{\rm cn})=(1-p_{\rm cn})\ketbra{\Psi_{\rm S}}+p_{\rm cn}\varrho_{\rm cn},
\eea
with $\ket{\Psi_{\rm S}}=(\ket{00}+\ket{11}+\ket{22})/\sqrt{3}$ and colored noise $\varrho_{\rm cn}=\mathbbm{1}_2\otimes \mathbbm{1}_2/4$ with $\mathbbm{1}_2=(\ket{0}\bra{0}+\ket{1}\bra{1})$. 

We add that we checked several random $3\times3$ edge states to test the conjecture of   Sanpera, Bru{\ss} and Lewenstein claiming that all bound entangled states in such systems have a Schmidt rank $2,$ and did not find a counter example \cite{SB01}.

{\it Evaluation of entanglement measures based on incomplete information.}---Experimentally it is very important that entanglement measures can be evaluated based on incomplete knowledge on the quantum state. There are efficient methods to bound  entanglement measures based on an operator expectation value from below \cite{GR08,EB07,WP09}. 
The current method can be adapted straightforwardly to the partial information case by replacing the condition $\omega_1 = \varrho$ with the set of linear constraints $\trace(\omega_1 O_i) = v_i,$ where $O_i$ are the measured observables and $v_i$ are the corresponding expectation values. 
As an example, see Fig.~\ref{elin_results}(b), where the entanglement is bounded
 from below based on complete information and 
based on $\exs{\sigma_x\otimes\sigma_x}$ and $\exs{\sigma_z\otimes\sigma_z}$ measurements for the state
\bea\label{partial}
\varrho_{\rm B}(p)=p_{\rm B} \ketbra{\Phi_{3\times 3}^+}+(1-p_{\rm B})\tfrac{1}{9}\openone,
\eea
where $\ket{\Phi_{3\times 3}^+}$ is a two-qubit Bell state $(\ket{00}+\ket{11})/\sqrt{2}$ embedded in the $3\times3$ system and $\sigma_l\otimes\sigma_l$
acts on this two-qubit system.

{\it Device independent scenario.}---The amount of entanglement can be bounded \emph{exclusively} from the observed data but independent of the quantum description. Depending whether only one or both sides are untrusted one distinguishes between a 
steering-type or a 
Bell-type scenario. 
The necessary steps to lift the method using only partial information to such device independent scenarios employs the translation idea highlighted in Ref.~\cite{MB14} and is explained in more detail in the supplement \cite{SUPP}. As an example, in Fig.~\ref{elin_results}(c) we plot a lower bound on the linear entropy of entanglement given as a function of the violation of the CHSH Bell inequality \cite{GT09}.

{\it Concave roof.}---Besides convex roofs, concave roofs can also be computed.
For example, if in Eq.~(\ref{Elininf1}) a concave roof is used instead of a convex roof, then we compute the 
linear entanglement of assistance \cite{DF99}, 
which is the maximal entanglement available if the mixed state is given as a purification to us, 
and a third party which holds the ancilla needed for the purifucation is assisting us.
In this case, in our method minimum must be replaced by maximum. 
In this way, we obtain a converging series of  {\it upper}
bounds on the entanglement of assistance. The results are shown in Fig.~\ref{elin_results}(d) for the family of $3\times3$ states 
\bea\label{state_for_Elin}
\varrho_E(p_{\rm E})=p_{\rm E}\ketbra{\Psi_{\rm E}}+(1-p_{\rm E})\tfrac{1}{9}\openone, 
\eea
where $\ket{\Psi_{\rm E}}=\epsilon \ket{00} +\epsilon \ket{11} + \sqrt{1-2\epsilon^2} \ket{22}$ and $\epsilon=0.3$.
As a reference, the linear entropy of entanglement is also shown for the same state.

{\it  Conclusions.}---We have shown a general framework for calculating convex roof-based entanglement measures. We demonstrated its use in calculating the entanglement for bipartite systems, as well as, the three-tangle for three-qubits. We discussed several other quantities for which it can be applied. 
In the future, we would like to explore further possibilities of using our algorithm to compute convex roofs, in calculating the linear Holevo capacity \cite{BW97,OV06}, the quantum Fisher information based on incomplete information \cite{qfi}, 
or the  convex or concave roofs of sums of several variances, as outlined in the Supplement  \cite{SUPP}.

We thank C. Eltschka, M. K\'us, J. Siewert, and M. Tiersch for stimulating discussions. We thank the
EU (ERC Starting Grant GEDENTQOPT, CHIST-ERA QUASAR, Marie Curie CIG 293993/ENFOQI),
the MINECO (Project No. FIS2012-36673-C03-03), the Basque
Government (Project No. IT4720-10), the OTKA (Contract No. K83858), 
the UPV/EHU program UFI 11/55,
the FQXi Fund (Silicon Valley Community
Foundation), and the DFG.

\eject

\renewcommand{\thefigure}{S\arabic{figure}}
\renewcommand{\thetable}{S\arabic{table}}
\renewcommand{\theequation}{S\arabic{equation}}
\setcounter{figure}{0}
\setcounter{table}{0}
\setcounter{equation}{0}
\newcommand{\loo}{{\lambda}}

\subsection{Supplemental Material}

In this supplemental material, we give some further details of our derivations. 

 \subsection* {Proof of Observation 2.}
 
 Let us assume that the state has $\elinprime(\varrho)=0.$ Then, from 
 Eq.~(\ref{Elinppt}) it follows that there is a symmetric PPT state $\omega_{12}$
such that ${\rm Tr}(\mathcal{A}_{AA'}\omega_{12})=0$ and ${\rm Tr_1}(\omega_{12})= \varrho.$ 
Hence, for this state
$\trace_{A'B'}(\omega_{12})=\varrho$ and
$\mathcal{F}_{AA'}\omega_{12}=\omega_{12}.$
The symmetry of $\omega_{12}$ means that $\mathcal{F}_{ AA'}\mathcal{F}_{BB'}\omega_{12}=\omega_{12}.$
Hence, $\mathcal{F}_{BB'}\omega_{12}=\omega_{12}$ also holds.
We can write
\begin{eqnarray}
\omega_{12}^{T_{AB}}=(\mathcal{F}_{AA'} \omega_{12} \mathcal{F}_{AA'})^{T_{AB}}
=\mathcal{F}_{AA'}\omega_{12}^{T_{A'B}} \mathcal{F}_{AA'}.
\end{eqnarray}
Since $\mathcal{F}_{AA'}$ is unitary, $\omega_{12}^{T_{AB}}\ge 0$ implies 
$\omega_{12}^{T_{A'B}}\ge 0.$ Finally, we obtain
\begin{eqnarray}
\varrho^{T_A}=\trace_{A'B'}( \omega_{12}^{T_{A'B}}).
\end{eqnarray} 
Hence,
$\omega_{12}^{T_{A'B}}\ge 0$ implies 
$\varrho^{T_{A}}\ge 0.$ 
This proves the first part of our observation.

To prove the second part, note that based on the discussion above
$\omega_{12}$ is a 2:2 symmetric extension of $\varrho.$ It is not necessarily a PPT symmetric extension
since for the $A:BA'B'$ partition it can also be non-PPT. 

Finally, the third part can be proved as follows. Let us assume that  $\varrho$ has a $2:2$ PPT symmetric extension denoted by $\omega_{12}.$
Hence, $\mathcal{F}_{AA'}\omega_{12}=\omega_{12}$ and
$\mathcal{F}_{BB'}\omega_{12}=\omega_{12}.$ Moreover, $\omega_{12}$ is a PPT state.
Hence, ${\rm Tr}(\mathcal{A}_{AA'}\omega_{12})=0.$ 

$\qed$

Note that Theorem 2 can be generalized to states that have $E_{\rm lin}^{({\rm n})}(\varrho)>0,$
involving PPT symmetric extensions and symmetric extensions to several parties.


 \subsection*{Quantitative entanglement witnesses}

In this section, we describe how our method can be used to construct  quantitative entanglement witnesses.
As an example, we present a condition for entanglement witnesses, such
that the expectation value of all witnesses satisfying the condition gives a lower 
bound on $E_{\rm lin}^{({\rm ppt})}$ defined in Eq.~\eqref{Elinppt}.
We also prove that for every state $\varrho_{AB}$ there is a witness of this type that gives not only a lower bound,
but gives the value of $E_{\rm lin}^{({\rm ppt})}$ exactly.

For the linear entropy of entanglement we needed to minimize the expectation value of the operator $\Coperator= \mathcal{A}_{AA^\prime} \otimes \mathbbm{1}_{BB^\prime}$ over all symmetric separable states $\omega_{12}$ with a fixed reduced marginal $\tr_2(\omega_{12})=\rho_{AB}$. Consider now an operator $W=W_{AB}$ that acts on the original bipartite Hilbert space. We require that is satisfies
\begin{equation}
\label{eq:witPPT}
\Coperator - \Pi_{\rm sym} (W \otimes \mathbbm{1}) \Pi_{\rm sym} =
P + \Pi_{\rm sym} Q^{T_1} \Pi_{\rm sym} 
\end{equation}
where $P,Q \geq 0$. Here $P$ is an operator acting only on the symmetric subspace of the two copies $Sym(\mathcal{H}_{AB}^{\otimes 2})$, while $Q$ acts on the full tensor product $\mathcal{H}_{AB}^{\otimes 2}$ but we only used the projected symmetric part of the partial transpose. For such a decomposition, it can be shown that its expectation value for $\omega_{12}$ is 
\begin{align}
\tr\{[&\Coperator - \Pi_{\rm sym}(W\otimes \mathbbm{1})\Pi_{\rm sym}] \omega_{12}\} \nonumber\\ 
&\;\;\;\;\;\;\;\;\;= \tr(P \omega_{12}) + \tr(\Pi_{\rm sym} \omega_{12} \Pi_{\rm sym} Q^{T_1})  \nonumber\\
&\;\;\;\;\;\;\;\;\;= \tr(P \omega_{12}) + \tr( \omega_{12} Q^{T_1})  \nonumber\\
&\;\;\;\;\;\;\;\;\;= \tr(P \omega_{12}) + \tr( \omega_{12}^{T_1} Q) \geq 0.\label{MPIW}
\end{align}
The projectors onto the symmetric subspace $\Pi_{\rm sym}$ can be dropped in the third line since $\omega_{12}$ is supported  only on it. In the last line we used $\tr(XY^{T_1})=\tr(X^{T_1}Y)$, while nonnegativity holds because all occuring operators are positive semidefinite. Hence, Eq.~(\ref{MPIW}) can be rewritten as
\begin{align}
\tr(\Coperator \omega_{12}) &\geq \tr(\Pi_{\rm sym}(W \otimes \mathbbm{1})\Pi_{\rm sym} \omega_{12}) \nonumber \\ &= \tr(W \rho_{AB}),
\label{MPIW2}
\end{align}
where we have further simplified the right-hand side using that $\omega_{12}$ has a fixed reduced density matrix. 
Since Eq.~(\ref{MPIW2}) holds for any valid state $\omega_{12}$, it holds in particular for the one yielding the linear entropy of entanglement, thus we arrive at
\begin{equation}\label{Wbound}
E_{\rm lin}(\rho_{AB}) \geq \tr(W \rho_{AB}).
\end{equation}
Hence the expectation value of our witness provides a lower bound
on the linear entropy of entanglement.

Next, we will show that for a given quantum state $\rho_{AB},$ if we optimize over 
all such witness operators, it is always possible to find one that 
saturates the inequality~(\ref{Wbound}).

\textbf{Observation 3.}---For the linear entropy of entanglement we obtain
\begin{equation}\label{Wsup}
E_{\rm lin}(\rho_{AB}) \geq E_{({\rm lin})}^{({\rm ppt})}(\rho_{AB}) = \sup_{W \in \mathcal{W}} \tr(W \rho_{AB}),
\end{equation}
with $\mathcal{W}$ being the set of all operators $W$ of Eq.~\eqref{eq:witPPT}. 


\textit{Proof.} The proof is given by applying the dual form of a semidefinite program~\cite{SDP}, which has been employed in a variety of different quantum information problems. In particular we refer to Ref.~\cite{PPTsymext} which explains such a procedure very nicely for the separability criterion based on symmetric extensions. We have structured the proof in two parts: In the first part, we show an equivalent formulation on the two-copy level. Afterwards we further simplify this dual problem to interpret it as an operator acting on a single density operator using techniques that were introduced in Ref.~\cite{PPTsymext}.

In the first part, we parse the original problem as given in Observation~$1$ into the form of a semidefinite program and invoke its dual, which provides the same solution. In order to achieve this one should note that the two conditions, $\omega_{12}$ just supported on the symmetric subspace and the linear equations $\tr_1(\omega) = \tr_2(\omega)=\rho_{AB}$ can be satisfied automatically with an appropriate ansatz $\omega_{12} ({\bf x}) = \omega_{12}^{\rm fix} + \sum_i x_i F_i$. Here $\omega_{12}^{\rm fix} = \sum_i s_i B_i$ is the fixed part of two-copy density operator such that the marginals equal to $\rho_{AB}$ (its precise form being discussed later), while the remaining part $\sum_i x_i F_i$ is the yet to be determined part on the symmetric subspace, \textit{i.e.}, the set of operators $\{ B_i\}_i \cup \{ F_i \}_i$ is a Hermitian operator basis for the symmetric subspace $Sym(\mathcal{H}^{\otimes 2})$. With this the  primal problem reads
\begin{eqnarray}
\inf&& \tr(\Coperator \omega_{12}^{\rm fix}) + \sum_i x_i \tr(\Coperator F_i) \\
\nonumber
\text{s.t.}&& \omega_{12} ({\bf x}) =\omega_{12}^{\rm fix} +  \sum_i x_i F_i \geq 0,   \\
\nonumber
&& \omega_{12} ({\bf x})^{T_1} =(\omega_{12}^{\rm fix})^{T_1} +  \sum_i x_i F^{T_1}_i \geq 0,
\end{eqnarray}
where one should note that $\omega_{12}({\bf x})$ acts on the symmetric subspace, while $\omega_{12} ({\bf x})^{T_1}$ acts on the full tensor $\mathcal{H}^{\otimes 2}$. 

Taking this into account, it is straightforward to invoke the dual and to derive an equivalent optimization problem. That this dual program provides the same solution is certified for instance via the Slater regularity condition~\cite{SDP}, which holds since this problem has an inner point, \textit{i.e.}, $\omega_{12} = \Pi_{\rm sym} (\rho_{AB} \otimes \rho_{AB})\Pi_{\rm sym} > 0$ if $\rho_{AB}$ has full rank; otherwise one should constrain $\mathcal{H}$ anyway to the range of $\rho_{AB}$. Since such reformulations have been carried out quite frequently, we refer here only to the literature, and continue with its solution, which is given by
\begin{eqnarray}
\label{eq:dual}
\sup&& \tr(Z_{\rm fix}  \omega_{12}^{\rm fix}) \\
\text{s.t.}&& \Coperator - Z_{\rm fix} = P_{\rm sym} + \Pi_{\rm sym} Q^{T_1} \Pi_{\rm sym}, \nonumber \\
&& Z_{\rm fix} = \sum_i z_i B_i, \nonumber\\
&&P_{\rm sym} \geq 0, Q \geq 0,\nonumber
\end{eqnarray}
where similarly $P_{\rm sym}$ acts on $Sym(\mathcal{H}^{\otimes 2})$ and $Q$ on the full tensor product space. This finishes the first part.

In the remaining part we show how the objective of Eq.~\eqref{eq:dual} can be interpreted as an operator on the single copy. For that we need some  structure of the fixed part $\omega_{12}^{\rm fix}$ that is given by the reduced state $\rho_{AB}$. The idea follows closely the ideas of Ref.~\cite{PPTsymext}, though we need to do it here for the symmetric subspace.

To start, note that any given density operator $\rho_{AB}$ can be written as $\rho_{AB}=\mathbbm{1}/d + \sum_i \trace(S_i \rho_{AB}) S_i$ with $\{S_i \}_i$ being an operator basis for the traceless Hermitian operators. Next let us define $O_i = \Pi_{\rm sym} ( S_i \otimes \mathbbm{1}) \Pi_{\rm sym}$. The expectation values of all these operators $O_i$ are completely determined by the reduced state $\tr(O_i \omega_{12}({\bf x})) = \tr(O_i \omega_{12}^{\rm fix}) = \tr(S_i \rho_{AB})$, and since all these state coefficients are independent this means that the set $\{ O_i \}_i$ is linearly independent. This implies a positive definite Gram matrix $G_{ij}=\tr(O_i O_j) > 0$, a unique inverse $G^{-1}$, and the existence of the operators $\tilde O_i = \sum_{j} (G^{-1})_{ij} O_j$. These operators are the corresponding orthogonal operators $\tr(\tilde O_i O_j ) = \delta_{ij}$, so that the fixed part becomes
\begin{align}
\omega_{12}^{\rm fix} &= \Pi_{\rm sym}/d_{\rm sym} + \sum_i \tr(O_i \omega_{12}^{\rm fix}) \tilde O_i \nonumber\\
&= \Pi_{\rm sym}/d_{\rm sym} + \sum_i \tr(S_i \rho_{AB}) \tilde O_i.
\end{align}
Note that also the desired dimensionality of $d^2$ matches, since $\tr_1(\omega_{12})=\tr_2(\omega_{12})=\rho_{AB}$ are excatly $d^2$ independent linear equations. To transfer this to the single copy level we write this solution in terms of a map applied to $\Lambda[\rho_{AB}]=\omega_{12}^{\rm fix}$,
\begin{equation}
\Lambda[\rho_{AB}] = \tr(\rho_{AB}) \Pi_{\rm sym}/d_{\rm sym} + \sum_i \tr(S_i \rho_{AB}) \tilde O_i.
\end{equation}
This map has the adjoint map, \textit{i.e.}, the map satisfying $\tr(X\Lambda[Y])=\tr(\Lambda^\dag[X]Y)$ for all matrices $X,Y$,
\begin{equation}
\Lambda^\dag[Z] = \tr(Z) \mathbbm{1}/d_{\rm sym} + \sum_i \tr(\tilde O_i Z) S_i.
\end{equation}
Via this we can finally make the connection to the single copy level by
\begin{align}
\tr(Z_{\rm fix} \omega_{12}^{\rm fix}) &= \tr(Z_{\rm fix} \Lambda[\rho_{AB}]) \nonumber \\ &= \tr(\Lambda^\dag[Z_{\rm fix}] \rho_{AB}) \equiv \tr(W_Z \rho_{AB}),
\end{align}
where we defined the single copy witness $W_Z=\Lambda^\dag[Z_{\rm fix}]$ in the last equation, parametrized in terms of the coefficients of $Z_{\rm fix}$. However since we want to have the witness as the open parameter we need to parametrize $Z_{\rm fix}(W)$ in terms of the witness $W$. Setting
\begin{align}
Z_{\rm fix}(W)&=\Pi_{\rm sym}(W\otimes \mathbbm{1})\Pi_{\rm sym}, \nonumber \\
&= d \tr(W) \Pi_{\rm sym} + \sum_i \tr(S_iW) O_i
\end{align}
achieves  $\Lambda[Z_{\rm fix}(W)]=W.$ Via this we can finally replace all occurrances of $Z_{\rm fix}$ in Eq.~\eqref{eq:dual} by $W$ and we obtain the stated result of the observation.
$\qed$

Note that one can obtain other quantitative entanglement witnesses if one replaces the decomposable structure, as given in Eq.~\eqref{eq:witPPT}, by a different entanglement witness condition. It is easy to see that if the operator $\Coperator-\Pi_{\rm sym}(W\otimes \mathbbm{1})\Pi_{\rm sym}$ is non-negative on separable states then $\tr(W\rho_{AB})$ gives a lower bound. 
Compared to other possibilities, the advantage of the witness~\eqref{eq:witPPT} is that the optimization~\eqref{Wsup} to get the lower bound can be carried out with
semidefinite programming.

We also add that if one only has measured a few observables $\{ O_i \}_i$ then to get a lower bound one merely has to add the constraint $W = \sum_i w_i  O_i,$ which means that the witness is a linear combination
of the measured observables with coefficients $w_i.$
Then, we have to optimize $\sum_i w_i v_i,$
where $v_i$ are the corresponding expectation values $\mean{O_i}_{\rho}.$

Finally, if one also wants quantitative entanglement witness for the other tasks one can proceed similarly. For instance, if one likes to bound the tangle one demands that $T-\Pi_{\rm sym}(W \otimes \mathbbm{1}^{\otimes 3}) \Pi_{\rm sym}$ is a non-negative on all fully separable states, thus it is an entanglement witness to test against full separability.

 \subsection*{Other quantities that can be calculated by our approach}
  
 {\it  Convex roof of the Meyer-Wallach measure.}---The Meyer-Wallach measure is an entanglement measure for pure states defined as \cite{MW03}
 \be \label{eq:MW1}
Q=\tfrac{1}{N}\sum_{n=1}^{N} 2S_{\rm lin}(\varrho_n),
\ee 
where $\varrho_n$ is the reduced state of the $n^{\rm th}$ qubits. This measure can be generalized to include
the reduced states of multi-qubit groups \cite{S04}.
Our method can calculate the convex roof of the measure \eqref{eq:MW1} and the generalized measures as well.

 {\it  Holevo capacity.}---The linear Holevo $\chi$ capacity is defined as  \cite{BW97,OV06}
  \be
 \chi_2(\Lambda)=\max_{\{p_{k},\ket{\Psi_k}\}} \left\{ S_{\rm lin}(\Lambda(\varrho))-\sum_k p_k S_{\rm lin}[\Lambda(\ket{\Psi_k})] \right\}.
 \ee
 It is a capacity measure for a channel $\Lambda.$ For qubit channels, explicit formula is
 given in Ref.~\cite{OV06}. 
   
{\it  Convex and concave roofs in entanglement conditions with the quantum Fisher information and the variance.}---
First, let us see simple entanglement conditions with the quantum Fisher information and the variance.
We start from the fact that for pure  $N$-qubit states
\be\label{JxyzN2}
\va{J_x}+\va{J_y}+\va{J_z}= \tfrac{N}{2}
\ee
holds.
Next, we need the fundamental properties of the quantum Fisher information $F_Q[\varrho,A]$  in our criteria \cite{qfi}:
(i) For pure states $F_Q[\varrho,A]$ equals four times the variance $\va{A}_{\varrho}.$ 
(ii) For mixed states, it is a convex function of the state.
Hence, for separable states follows  \cite{T12}
\be\label{FJxJyJz}
\tfrac{1}{4}\sum_{l=x,y,z} F_Q[\varrho,J_l]\le \tfrac{N}{2}.
\ee
Due to the concavity of the variance, we can obtain a similar entanglement condition with variances as  \cite{TK07}
\be\label{varJxJyJz}
\va{J_x}+\va{J_y}+\va{J_z}\ge \tfrac{N}{2}.
\ee
Any state that violates Eq.~\eqref{FJxJyJz} or Eq.~\eqref{varJxJyJz} is entangled.

\begin{figure*}
\centerline{ \epsfxsize3.4in  \epsffile{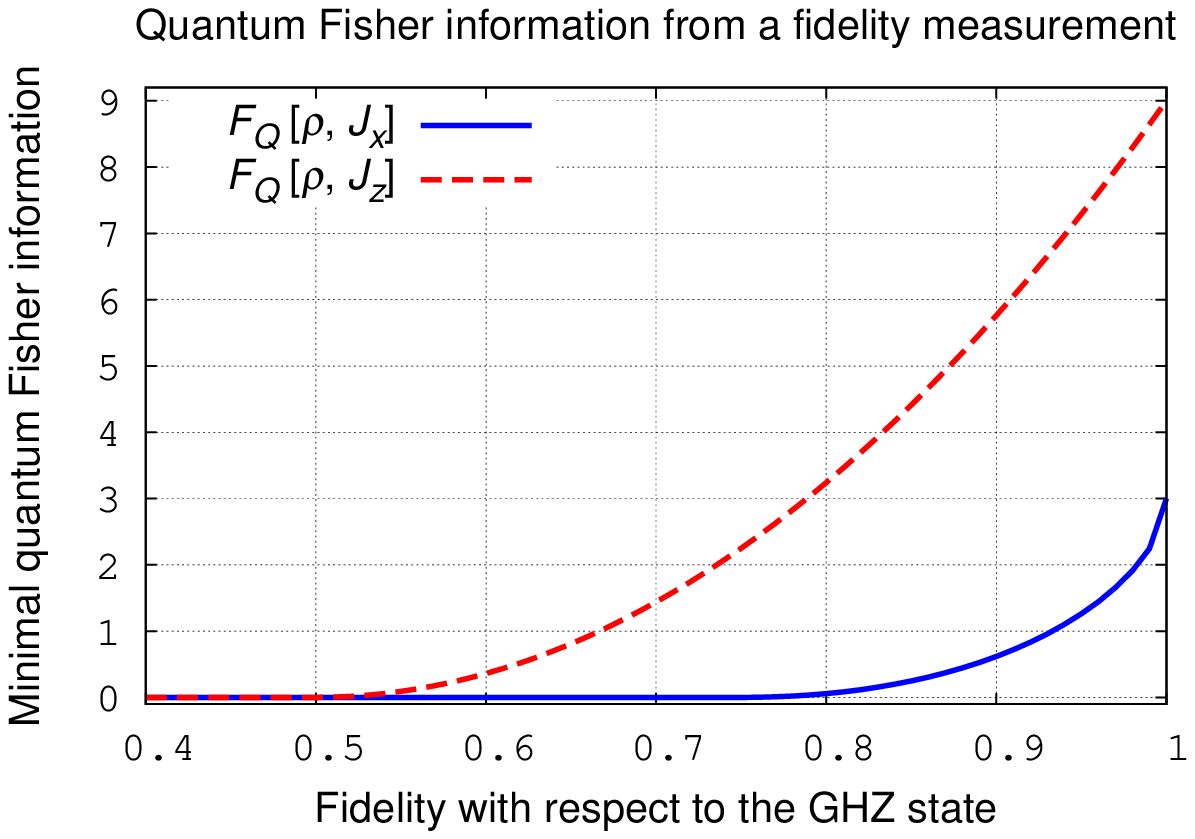}  \epsfxsize3.4in  \epsffile{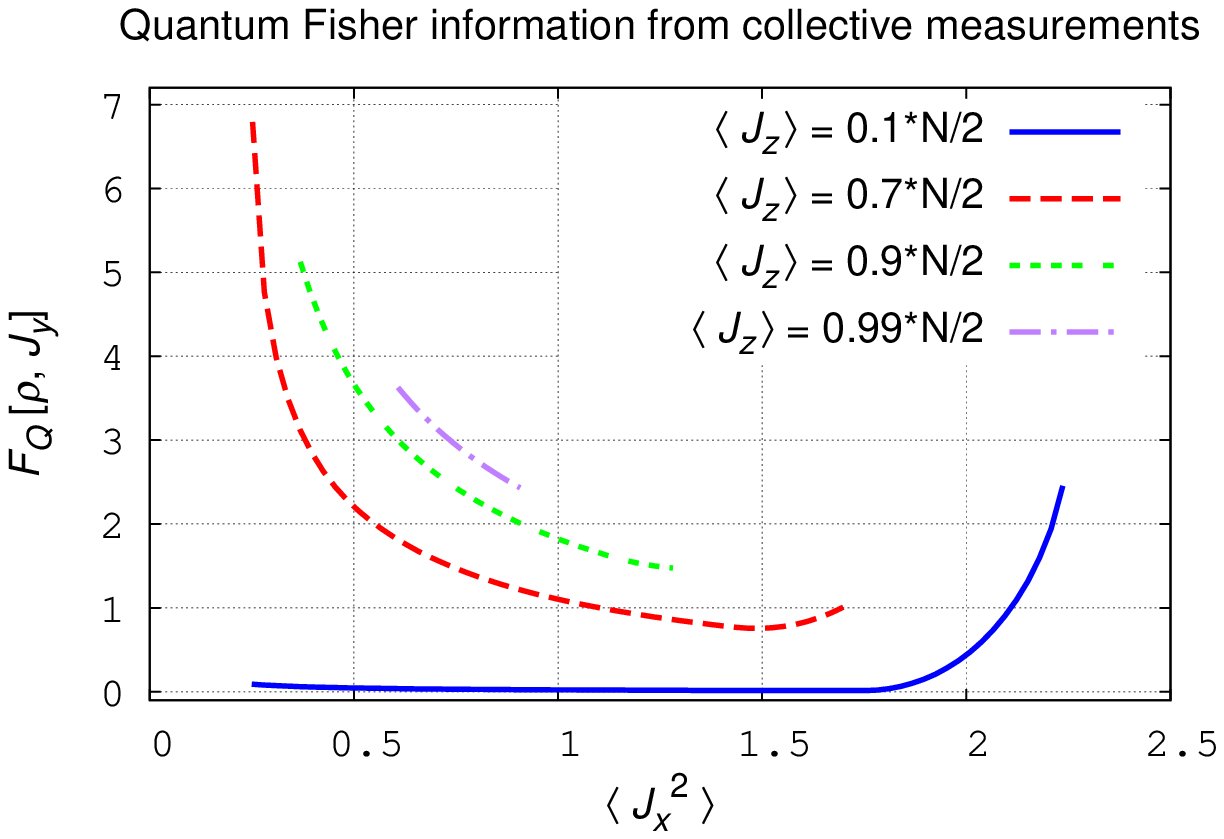}}
\vskip0.2cm
\centerline{ (a) \hskip8cm (b)}
\vskip0.0cm
\caption{
(a) Lower bound on $F_Q[\varrho,J_z]$ 
based on the fidelity with respect to the three-qubit GHZ state.
(b) Lower bound on $F_Q[\varrho,J_y]$ 
based on $\exs{J_x^2}$ for various values of  $\exs{J_z}$ and for $\exs{J_x}=0,$ for $N=3$ qubits.}
\label{figs:fisher}
\end{figure*}

The conditions \eqref{FJxJyJz} and \eqref{varJxJyJz} can be improved if we take the concave and convex roofs, respectively, 
of the left-hand sides of Eq.~\eqref{JxyzN2}.
Hence, alternative separability conditions arise
\be\label{infJxJyJz}
\min_{\{p_{k},\ket{\Psi_k}\}}\sum_{k}p_{k} \sum_{l=x,y,z} \va{J_l}_{\Psi_k}\le \tfrac{N}{2},
\ee
and 
\be\label{supJxJyJz}
\max_{\{p_{k},\ket{\Psi_k}\}}\sum_{k}p_{k} \sum_{l=x,y,z} \va{J_l}_{\Psi_k}\ge \tfrac{N}{2}.
\ee
Any state that violates these is entangled.
Numerical evidence shows that 
Eq.~(\ref{infJxJyJz}) is stronger than Eq.~(\ref{FJxJyJz}).
Moreover, numerical evidence shows also that Eq.~(\ref{supJxJyJz}) is stronger than Eq.~(\ref{varJxJyJz}).
These ideas can be extended to improve other entanglement conditions based on variances \cite{G04}.

We note that Ref.~\cite{LP13} shows that $2\times2$ covariance matrices $C_\varrho(A,B)$ 
can always be decomposed as the 
\be
C_\varrho(A,B)=\sum_k p_k C_{\Psi_k}(A,B),
\ee
where $\varrho$ has the decomposition as in Eq.~(\ref{Elininf}).
Hence, we know that the bound on the sum of two variances cannot be improved this way.
However, Ref.~\cite{VP13} demonstrates that such a decomposition is not always possible for
$3\times3$ covariance matrices. This is connected to the fact that the bound for separable states
for the sum of three variances can be improved.

{\it Quantum Fisher information based on incomplete data.}---
The quantum Fisher information can be bounded from below from partially known data.
That is, we know the expectation value of some operators, and want to
find a lower bound for the quantum Fisher information. The problem can be mapped
to a semidefinite optimization
in the two-copy space. A very good lower
bound can be obtained if we optimize over
PPT states.

For that we can use that the quantum Fisher information is, apart from a constant factor, the convex roof of the variance \cite{TP13}
\be
F_Q[\varrho,A]=4\min_{\{p_k,\ket{\Psi_k}\}} \sum_k p_k \va{A}_{\Psi_k}.
\ee
The variance of a pure  state $\ket{\Psi}$ can be expressed on two copies as
\be
\va{A}_\Psi=\trace[(A^2\otimes \openone-A\otimes A)\ketbra{\Psi}\otimes \ketbra{\Psi}].
\ee
Hence, a lower bound on the quantum Fisher information can be obtained as
\begin{eqnarray}
F_Q^{({\rm ppt})}[\varrho,A]=\nonumber
\min_{ \omega_{12} }&&
\trace[(A^2\otimes \openone-A\otimes A)\omega_{12} ],\\
\textrm{s.t.}&&
\omegasppt, \nonumber\\
&&\trace(O_i \omega_{1}) =v_i,\label{Fqopt}
\end{eqnarray}
where the constraints are given with the expectation values $v_i=\exs{O_i}_\varrho.$
The optimization \eqref{Fqopt} can straightforwardly be carried out
with semidefinite programming.

In Fig. \ref{figs:fisher}(a), we present a simple example where a lower
bound on the quantum Fisher information $F_Q[\varrho,J_z]$ is shown
based on measurements of the fidelity with respect to the GHZ state.
Below a fidelity of $\ifrac{1}{2},$ the bound for $F_Q[\varrho,J_z]$ is zero. This is due to the fact that the product state $\ket{11..111}$ reaches this fidelity value,
while $F_Q[\varrho,J_z]$ is zero for this state.
If the fidelity is $1,$ we obtain $F_Q[\varrho,J_z]=N^2$ and  $F_Q[\varrho,J_x]=N,$
as expected \cite{T12}.

In Fig. \ref{figs:fisher}(b), we present a bound on the quantum Fisher information
based on collective measurements, relevant to spin squeezing.
Note that for well polarized ensembles, increasing $\exs{J_x^2}$
leads to decreasing $F_Q[\varrho,J_y].$ On the other hand,
for small $\exs{J_z},$ increasing $\exs{J_x^2}$
leads to increasing $F_Q[\varrho,J_y].$
Some of the curves have points only
in certain ranges of $\exs{J_x^2},$
as there are no physical states
corresponding to measurement results outside of these ranges,
assuming a given value for $\exs{J_z}$ and $\exs{J_x}=0.$

Similar methods can be used for bounding the variance
of an observable from above based on the expectation value of
other observables. 
We can use that the variance is the concave roof of itself \cite{TP13}
\be
\va{A}=\max_{\{p_k,\ket{\Psi_k}\}} \sum_k p_k \va{A}_{\Psi_k}.
\ee
The difference between the two
cases is that for the quantum Fisher information
we have to look for the minimum, while for the variance
we have to look for the maximum. 

 {\it  Genuine multipartite entanglement.}---It is possible to define quantities that detect true multipartite entanglement and can be evaluated with our method.
Let us define 
 \be \label{GG} G=\min_{\{p_k,\ket{\Psi_{k}}\}}
 \sum_{k}p_{k}\prod_n S_{{\rm lin}n}(\ket{\Psi}_k)
 \ee 
 where $S_{{\rm lin}n}(\ket{\Psi})$ is the linear entropy for the $n^{\rm th}$  bipartition of the qudits.
 To be more precise,  $S_{{\rm lin}n}(\ket{\Psi})$ is the linear entropy of the reduced state
 of the qudits in one of the two partitions for the $n^{\rm th}$  bipartion.
 If $G=0$ then the state is biseparable, otherwise it is genuine multipartite entangled.

 Similar idea can work such that only a sum of entropies must be computed by defining
 \begin{eqnarray}\label{HH}
H=
\min_{ \{\varrho_n\} }&&
\sum_n
p_n E_{{\rm lin}n}(\varrho_n),\\
\textrm{s.t.}&&
\sum p_n \varrho_n=\varrho, \nonumber
\end{eqnarray}
  where $E_{{\rm lin}n}$ is linear entropy for the $n^{\rm th}$ bipartition.
  If $H=0$ then the state is biseparable, otherwise it is genuine multipartite entangled.
  If, instead of $E_{{\rm lin}}(\varrho)$, we calculate $E_{{\rm lin}}^{({\rm ppt})}$ given in Eq.~\eqref{Elinppt}
  then  Eq.~\eqref{HH} can be obtained via a semidefinite program.
  The advantage of Eq.~\eqref{HH} is that only two copies of the original state are needed to calculate the value with our approach,
  while for the formula \eqref{GG} we need much more copies.
 The formalism of  Eq.~\eqref{HH} is in the spirit of the PPT mixer detecting genuine multipartite entanglement \cite{PPTmixer}.
  
Note that a three-qubit state mixed from states that are PPT with respect to some partitions have been found 
that is genuine multipartite entangled \cite{Tobi}. 
Thus, detecting genuine multipartite entanglement is a non-trivial task.

\subsection{Device independent programs}

In this section, we explain the methods to obtain lower bounds on the linear entropy of entanglement for the device independent scenarios; either in the steering case where only the apparatus of one side is uncharacterized, or in Bell-type scenarios where both sides are unknown.

We will use the tool presented in Ref.~\cite{MB14}, resting on ideas from Ref.~\cite{NPA1,doherty08}, which transforms the problem of estimating entanglement in a device independent scenario into the more common problem to lower bound the entanglement of a given fixed finite-dimensional system having only partial information. The method uses instead of the quantum state $\varrho$ of unkown dimension, a finite dimensional object $\chi$ which captures most of the properties of the state.

To set the stage, let us assume that on a given side, say system $A$, one only knows the number of settings $x=1,\dots,n$ and respective outcomes $a=1,\dots,m$. This measurement scheme is described by a collection of POVM elements $M_{a|x}$, which act on a Hilbert space $\mathcal{H}_A$ of unknown dimension. To this measurement scenario one now associates a specific completely positive local map: $\Lambda_A(\varrho_A) = \sum_k F_k \varrho_A F_k^\dag$ with Kraus operators $F_k = \sum_{\vec i} \ket{ {\bf i}}_{\bar A\:A} \bra{k}M_{\bf i}$. Here $\ket{k}_A$ and $\ket{{\bf i}}_{\bar A}$ are respective basis states of the input and output Hilbert spaces, while $M_{\bf i} $ are operators out of a chosen set $\mathcal{M}$ on which we comment shortly. However, via this structure, first observe that this map transforms a given input state $\varrho_A$ to 
\begin{equation}
\chi_{\bar A} =\Lambda_A[\varrho_A]= \sum_{\bf i,j}\ket{{\bf i}}\bra{{\bf j}} \tr(M_{\bf j}^\dag M_{\bf i} \varrho_A),
\end{equation}
hence an output with matrix elements given by certain expectation values. At this stage the specific operator set $\mathcal{M}$ becomes important, since so far we know nothing about $\chi_{\bar A}$ because we neither know $\varrho_A$ nor $M_{\bf i}$. The only knowledge that we have are certain generic properties of the POVM elements $M_{a|x}$, more precisely we have $(i)$ positivity $M_{a|x}\geq 0$, $(ii)$ normalization $\sum_a M_{a|x}=\mathbbm{1}$ and $(iii)$ that each operator $M_{a|x}$ is a projector.  Here note that by Naimark's extension any measurement can be written
as a projector onto a larger dimensional space. Since for most device independent tasks this extension does not change the underlying tasks this property can be assumed without loss of generality. $M_{a|x}M_{a^\prime|x}=\delta_{aa^\prime} M_{a|x}$. In addition note that the expectation values of each measurement operator is observable, $(iv)$ $\tr(M_{a|x}\varrho_A)=P(a|x)$. 

Via these four properties one can thus choose specific operator sets $\mathcal{M}$ such that one has at least some partial information on $\chi_{\bar A}$. For instance, if one chooses $\mathcal{M}$ to consist of the measurement operators $\mathcal{M}=\{ M_{a|x} \}_{a,x}$ one knows for instance 
\begin{eqnarray}
\tr(M_{a|x} M_{a^\prime|x} \varrho_A) &=& \delta_{aa^\prime}\tr(M_{a|x} \varrho_A) \nonumber\\
&=& \delta_{aa^\prime} P(a|x),
\end{eqnarray}
while other entries like $\tr(M_{a|x} M_{a^\prime|x^\prime} \varrho_A)$ with $x\not = x^\prime$ are still unknown. Nevertheless via this one gets some partial knowledge and some structure of $\chi_{\bar A}$, which can be captured by an explicit parametrization as
\begin{align}
\chi_{\bar A}[P,u] &= \chi^{\rm fix}[P] + \chi^{\rm open}[u] \nonumber\\
&= \sum_{a|x} P(a|x) Z_{a|x} + \sum_v u_v Z_v, 
\end{align}
using appropriate operators $Z_{a|x}$ and $Z_v$.
Here the first part represents the known part of $\chi_{\bar A}$, while the second one is the restricted open unknown part.

Such a structure can be inferred for any choice of $\mathcal{M}$. For instance, one could remove some linear dependencies of the just given example set if one adds the identity and erases the last outcome for each measurement setting $\mathcal{M}_1=\{ \mathbbm{1}\} \cup \{ M_{a|x}\}_{a<n,x}$. In addition note that one could also enlarge this set by including also products up to $N$ POVM elements $\mathcal{M}_N$, so for instance $\mathcal{M}_2= \mathcal{M}_1 \cup \{ M_{a|x} M_{a^\prime|x^\prime}\}_{a,a^\prime < n, x\not =x^\prime}$, already removing trivial parts. In this way one gets further relations like
\begin{equation}
\sum_{a^\prime} \tr(M_{a|x} M_{a^\prime|x^\prime} M_{a^\prime|x^\prime} M_{a|x}) = P(a|x)
\end{equation}
if $x\not = x^\prime$. The advantage of including products is that one gets a tighter, more constrained, description. This set of operators $\mathcal{M}_N$ is precisely the one which has been mostly used~\cite{MB14}, since it is very straightforward to ``decode'' all the known structure. Still there are other possibilities, like $\mathcal{M}_t= \{ M_{a_1|1} M_{a_2|2} \dots M_{a_n|n} \}_{a_1,\dots,a_n}$. Here it might be harder to deduce all the structure but it has for instance the advantage that the associated map $\Lambda_A$ is then even trace-preserving, thus $\chi_{\bar A}$ can be completely interpreted as an output quantum state; something which is not directly possible if one uses $\mathcal{M}_N$.

Now let us come to the concrete cases. At first let us discuss the fully device independent case where both sides are completely uncharacterized. If we locally apply the just described trace-preserving physical map (using for instance the choice $\mathcal{M}_t$) we transform any state $\varrho_{AB}$ into another bipartite state $\chi_{\bar A \bar B}=\Lambda_{A} \otimes \Lambda_B(\varrho_{AB})$. Since an entanglement monotone does not increase under local operators and classical communication, we get $E(\varrho_{AB}) \geq E(\chi_{\bar A \bar B})$ and thus we obtain a valid lower bound by estimating the entanglement of the output state. Hence if we want to bound the linear entropy of entanglement by seeing a certain value of a Bell inequality $I \cdot P = V$ we use
\begin{eqnarray}
E_{\rm min}(I\cdot P=V) \\
\geq \min_{\omega_{12},u,P}&& \tr( \mathcal{A}_{12} \omega_{12})\nonumber\\
\nonumber
\textrm{s.t.}&& \omega_{12} \text{ is $d_{\bar A} \times d_{\bar B},$ symmetric, separable,}\\
\nonumber 
&& \omega_1=\chi_{\bar A \bar B} = \chi_{\bar A \bar B}^{\rm fix}[P] + \chi_{\bar A \bar B}^{\rm open}[u] \geq 0, \\
\nonumber 
&& I_{\rm chsh} \cdot P = V,
\end{eqnarray}

Now let us turn to the steering case, where we assume that Alice's side is uncharacterized while Bob obtains complete tomography. Then the data are  given by the collection of unnormalized density operators $\mathcal{E}=\{ \varrho_{a|x} \}_{a,x}$ for Bob with $P(a|x)=\tr(\varrho_{a|x})$. In principle we can use the same method as for the fully device independent case by employ the trace-preserving local map only on one side $\chi_{\bar A B} = \Lambda_{A} \otimes \rm{id}[\varrho_{AB}]$ and then bounding the linear entropy of entanglement of the output state. 

However in this case we can do slightly better, since it is possible to bound the linear entropy of entanglement more directly on the original state $\varrho_{AB}$. This is in similar spirit as the negativity of Ref.~\cite{MB14} and Ref.~\cite{pusey}. Suppose we apply the same local, not necessarily trace-preserving, local map to the two copies $\chi_{12} \equiv\chi_{\bar A \bar A^\prime B B^\prime}=\Lambda_{A} \otimes \Lambda_{A} \otimes \rm{id} [\omega_{12}]$. Then we can relax the constraint that
$\omega_{12}$ is the symmetric PPT state of two qudits 
by 
$\chi_{12} \geq 0$, $\chi_{12}^{T_1}\geq 0$ and the permutation invariance \begin{eqnarray}
\mathcal{F} \chi_{12} \mathcal{F}\! & =&  \Lambda_{A} \otimes \Lambda_{A} \otimes \textrm{id} [\mathcal{F} \omega_{12} \mathcal{F}]\! \nonumber\\
& =&  \Lambda_{A} \otimes \Lambda_{A} \otimes \rm{id} [\omega_{12}] =\chi_{12}.
\end{eqnarray}
Note that since the identity is within the set $\mathcal{M}$ we have that the data of $\chi_{AB}=\Lambda_{A} \otimes \textrm{id}[\varrho_{AB}]$ are included in $\chi_{12}$, thus we can directly parametrize $\chi_{12} = \chi_{12}^{\rm fix}[\mathcal{E}] + \chi_{12}^{\rm open} [u]$. 
The key difference compared to the previous case is that the objective value is still accessible. 
Due to the symmetry of the linear entropy of pure states, it is not surprising that  $\mathcal{A}_{AA^\prime}\mathbbm{1}_{BB^\prime}=\mathbbm{1}_{AA^\prime}\mathcal{A}_{BB^\prime}$ holds on the symmetric subspace of the two copies. However, because the identity is included in $\mathcal{M}$ and because Bob's side is charaterized, 
the expectation value of $\mathbbm{1}_{AA^\prime}\mathcal{A}_{BB^\prime}$ is given by a linear function of the values $\chi_{12}$; this linear function is denoted by  $\tr(\hat E\chi_{12})$ as a shorthand.
Then we get as a lower bound 
\begin{eqnarray}
\label{eq:Elinsteer}
E_{\rm min}(\mathcal{E}) \geq \min_{u} && \tr(\hat E\chi_{12}) \\
\nonumber
\textrm{s.t.} &&  \chi_{12}=\chi_{12}^{\rm fix}[\mathcal{E}] + \chi_{12}^{\rm open}[u] \geq 0,\: \chi_{12}^{T_1} \geq 0.
\end{eqnarray}

Before we conclude, let us point out that for both programs one can obtain sharper bounds if one includes more products into the generating set $\mathcal{M}$. This is very straightforward for the steering case, but even such programs quickly reach the border of being feasible. Here it remains to investigate which particular sets $\mathcal{M}$ are more suitable than others. We leave this open for further investigation. If one combines these ideas with, for instance, the Schmidt number program then one could access even the Schmidt number (often taken as a synonym for quantum dimension) also in a device independent way.

\begin{figure}[t]
\vskip0.5cm
\centerline{ \epsfxsize3.4in  \epsffile{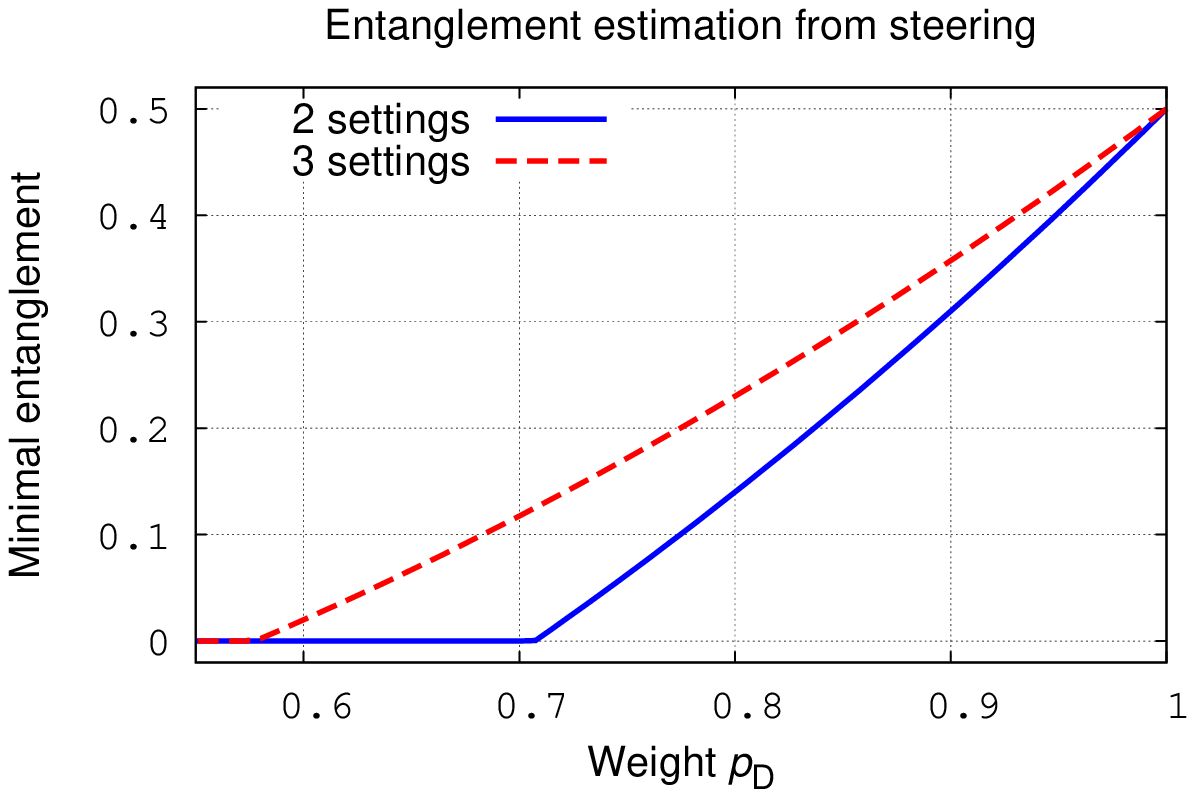}}
\vskip-0.5cm
\caption{
Entanglement quantification via the simplest steering cases according to Eq.~\eqref{eq:Elinsteer2}.}
\label{figs:steering}
\end{figure}

As an example we were investigating the simplest steering scenarios, where Alice, the uncharacterized side, has either two or three dichotomic measurements and Bob performs full tomography on his qubit. 
 In this case the available data are given by the corresponding probabilities for Alice $P(r|s)$, with $r \in \{\pm 1\}$ denoting the outcome and $s \in \{1,2,3\}$ the setting, and the corresponding conditional states for Bob $\rho_{r|s}$. 

For the examples we assume the following data
\begin{equation}
\label{eq:Elinsteer2}
P(r|s)=\tfrac{1}{2}, \;\;\rho_{r|s}=\tfrac{1}{2}(\mathbbm{1} - r p_{\rm D} \sigma_s ), 
\end{equation}
with $\sigma_s$ denoting the standard Pauli matrices. Such data are for instance generated by measuring the noisy singlet state of two qubits in the standard spin directions. Figure~\ref{figs:steering} shows the corresponding lower bound for the linear entropy of entanglement. We remark that the points where the lower bound becomes trivial coincide with the analytic cut-off value $p_{\rm D} \leq 1/\sqrt{2}$ and $p_{\rm D} \leq 1/\sqrt{3}$ respectively, while the lower bound at the maximum, $E_{\rm min} = 1$ is exact, since such observations require at least a maximally entangled two-qubit state.


\begin{thebibliography}{References}

%
%
\bibitem{GT09}R. Horodecki  {\it et al.}, 
Rev. Mod. Phys. \textbf{81}, 865 (2009); O. G\"uhne and G. T\'oth,
Phys. Rep. \textbf{474}, 1 (2009).

%
%
\bibitem{expdet}J.-W. Pan {\it et al.}, 
Nature (London) \textbf{403}, 515 (2000);
M.~Bourennane {\it et al.}, 
Phys. Rev. Lett. \textbf{92}, 087902 (2004);
W. Wieczorek {\it et al.}, 
Phys. Rev. Lett. {\bf 103}, 020504 (2009).

%
%
\bibitem{GR08}  O.~G\"uhne, M.~Reimpell, and R.~F.~Werner
Phys. Rev. Lett. {\bf  77}, 052317 (2008). 
\bibitem{EB07}  J. Eisert, F. G. S. L. Brandao, and K. M. R. Audenaert,
New J. Phys. {\bf 9}, 46 (2007).
\bibitem{WP09} H.~Wunderlich and M.~B. Plenio, J. Mod. Opt. {\bf 56}, 2100 (2009).

%
%
\bibitem{OA02}  A. Osterloh {\it et al.}, 
Nature (London) {\bf 416}, 608 (2002);
M.~Cramer {\it et al.},
Nat. Commun. {\bf 4}, 2161 (2013).

%
%
\bibitem{WK02}   W.~K.~Wootters, Phys. Rev. Lett. {\bf 80} 2245, (1998).
\bibitem{WG03} T.-C.~Wei and P.~M.~Goldbart, Phys. Rev. A {\bf 68}, 042307 (2003).
\bibitem{CK00} V.~Coffman, J.~Kundu, and W.~K.~Wootters, Phys. Rev. A {\bf 61}, 052306 (2000).
\bibitem{AB01} A. ~Ac\'{\i}n  {\it et al.}, 
Phys. Rev. Lett. \textbf{87}, 040401 (2001).
\bibitem{SB01} A.~Sanpera, D.~Bru\ss, and  M.~Lewenstein, 
Phys. Rev. A {\bf 63}, 050301 (2001).
\bibitem{RL09} B.~R\"othlisberger, J.~Lehmann, and D.~Loss, Phys. Rev. A {\bf 80}, 042301 (2009).

%
%
\bibitem{VMMW02} F. Verstraete and M.~M. Wolf, Phys. Rev. Lett. {\bf 89}, 170401
(2002).
\bibitem{LV11} Y.-C.~Liang, T.~V\'ertesi, and N.~Brunner, Phys. Rev. A {\bf 83},
022108 (2011).
\bibitem{MB14}  T.~Moroder {\it et al.}, 
Phys. Rev. Lett. {\bf 111}, 030501 (2013).
\bibitem{pusey} M.~Pusey,  Phys. Rev. A {\bf 88}, 032313 (2013) 

%
%
\bibitem{Itangle}  
P.~Rungta {\it et al.}, 
Phys. Rev. A {\bf 64}, 042315 (2001).

\bibitem{elinmonotone} 
P. Rungta and C. M. Caves,
Phys. Rev. A {\bf 67}, 012307 (2003);
G. Vidal, J. Mod. Opt.  {\bf 47}, 355 (2000).
\bibitem{ElinEnt} M. Huber, and J. I. de Vicente, 
Phys. Rev. Lett. {\bf 110}, 030501 (2013);
M. Huber, M. Perarnau-Llobet, and J.I. de Vicente,
Phys. Rev. A {\bf 88}, 042328 (2013).

%
%
\bibitem{H03} P. Horodecki, Phys. Rev. A {\bf 68}, 052101 (2003).


\bibitem{MK05}
F.~Mintert, M.~Kus, and A.~Buchleitner,
{Phys. Rev. Lett.} {\bf 95}, 260502 (2005).

%
%
\bibitem{W89} R.~F. Werner, Phys. Rev. A {\bf 40}, 4277 (1989).
\bibitem{SymSep} Every symmetric separable state of $N$ particles can be written as  $\sum_k q_k \vert \Phi_k \rangle \langle \Phi_k\vert ^{\otimes N},$ $\sum_k q_k=1,$ and $q_k>0.$
See J.~Korbicz, J.~I. Cirac, and M.~Lewenstein, Phys. Rev. Lett. {\bf 95}, 120502 (2005).

%
%
\bibitem{PPT} A.~Peres, Phys.  Rev. Lett.  {\bf 77}, 1413 (1996); M.~Horodecki, P.~Horodecki,
and R.~Horodecki, Phys. Lett. A  {\bf 223}, 1 (1996).


%
%
\bibitem{bound}
M.~Horodecki {\it et al.}, 
Phys. Rev. Lett. {\bf 80}, 5239 (1998);
for a review see P. Horodecki in D. Bru{\ss} and
G.~Leuchs (eds.), {\it Lectures on Quantum Information}
(Wiley-VCH, Berlin, 2006).
\bibitem{H97} P. Horodecki, Phys. Lett. A {\bf 232}, 333 (1997).
\bibitem{boundinteresting}
K.~Horodecki {\it et al.},
Phys. Rev. Lett. {\bf 94}, 160502 (2005);
A.~Ac\'{\i}n {\it et al.},
Phys. Rev. Lett. {\bf 92}, 107903 (2004);
G.~T\'oth {\it et al.},
Phys. Rev. Lett. {\bf 99}, 250405 (2007).


%
%
\bibitem{CM12} Z.-H.~Chen, Z.-H.~Ma, O.~G\"uhne, and S.~Severini, Phys. Rev. Lett. {\bf 109}, 200503 (2012).



\bibitem{mathworks} The CoRoNa package can be downloaded at 
\href{http://www.mathworks.com/matlabcentral/fileexchange/47823-corona-convex-roof-numerical-analysis}{http://www.mathworks.com/matlabcentral/fileexchange/47823-corona-convex-roof-numerical-analysis}.


%
%
\bibitem{CA05}
K. Chen, S. Albeverio, and S.-M. Fei,
{Phys. Rev. Lett.} {\bf 95}, 040504 (2005);
J. I. de Vicente,
Phys. Rev. A {\bf 75}, 052320 (2007); {\bf 77}, 039903(E) (2008);
O.~Gittsovich and O.~G\"uhne,
Phys. Rev. A {\bf 81}, 032333 (2010).
\bibitem{G03}
E.~Gerjuoy,
Phys. Rev. A {\bf 67}, 052308 (2003);
Y.-C. Ou, H. Fan, and S.-M. Fei,
Phys. Rev. A {\bf 78}, 012311 (2008);
M.~Li, S.-M.~Fei, and Z.-X.~Wang,
J. Phys. A: Math. Theor. {\bf 42}, 145303 (2009).

\bibitem{PPTsymext} A.~C.~Doherty, P.~A.~Parrilo, and F.~M.~Spedalieri, Phys. Rev. A {\bf 69}, 022308 (2004); {\bf 71}, 032333 (2005).

%
%
\bibitem{SUPP} See Supplemental Material, which includes
Refs.~\cite{MW03,S04,T12,TK07,G04,TP13,VP13,LP13,NPA1,doherty08,Tobi}.
\bibitem{MW03} D.~A.~Meyer and N.~R.~Wallach, J. Math. Phys. {\bf 43} 
4273 (2002). 
\bibitem{S04} A.~J.~Scott, Phys. Rev. A {\bf 69}, 052330 (2004). 
\bibitem{T12} G.~T\'oth, Phys. Rev. A {\bf 85}, 022322 (2012);
P.~Hyllus {\it et al.},
Phys. Rev. A  {\bf 85}, 022321 (2012).
\bibitem{TK07} G.~T\'oth {\it et al.}, 
Phys. Rev. Lett. {\bf 99}, 250405 (2007).
\bibitem{G04} O.~G\"uhne, Phys. Rev. Lett. {\bf 92}, 117903 (2004).
\bibitem{LP13} Z.~L\'eka and D.~Petz,  Probab. Math. Statist.  {\bf 33}, 191 (2013).
\bibitem{VP13} D.~Virosztek and D.~Petz,   arXiv:1311.3908.
\bibitem{TP13} G.~T\'oth and D.~Petz, Phys. Rev. A {\bf 87}, 032324 (2013);
S. Yu, arXiv:1302.5311.
\bibitem{PPTmixer} B. Jungnitsch, T. Moroder, and O. G\"uhne, 
Phys. Rev. Lett. {\bf 106}, 190502 (2011).
\bibitem{Tobi} T. Moroder and O. G\"uhne, unpublished (2014). This problem is in the problem book of the National Quantum Information Centre, Gdansk.
\bibitem{NPA1} M. Navascu\'es, S. Pironio, A.~Ac\'{\i}n, Phys. Rev. Lett. {\bf 98}, 010401 (2007);
New J. Phys. {\bf 10}, 073013 (2008) 
\bibitem{doherty08} A.~C.~Doherty {\it et al.}, 
Proceedings of IEEE Conference on Computational Complexity 2008, pages 199--210; arxiv:0803.4373.

%
%
\bibitem{lur} O. G\"uhne,  Phys. Rev. Lett. {\bf 92}, 117903 (2004).
\bibitem{GH07} O. G\"uhne {\it et al.}, 
Phys. Rev. Lett. {\bf 99}, 130504 (2007).
\bibitem{ccnr}
O.~Rudolph,
Quantum Inf. Process. {\bf 4}, 219 (2005); K. Chen and L.-A. Wu, Quantum Inf. Comput. {\bf 3}, 193 (2003).

%
%
\bibitem{TG09}
G.~T\'oth and O.~G\"uhne, Phys. Rev. Lett. {\bf 102}, 170503 (2009);
Appl. Phys. B {\bf 98}, 617 (2010).

\bibitem{symext} This extension is different from the one in Observation 2 as it adds new copies of the entire bipartite system, rather than adding new copies of the parties A and B. Moreover, it can be shown that since the state $\omega_{12}$ is symmetric, 
an $1:n$ PPT symmetric extension is the same as an $1+\Delta:n-\Delta$ extension for any integer $0<\Delta<n.$

%
%
\bibitem{NO09} M.~Navascu\'es, M.~Owari, and M.~B.~Plenio, Phys. Rev. A {\bf 80}, 052306 (2009).

\bibitem{OS12} A.~Osterloh and J.~Siewert, Phys. Rev. A {\bf 86}, 042302 (2012).
\bibitem{G05} G. Gour, Phys. Rev. A {\bf 71}, 012318 (2005).


%
%
\bibitem{DV00} W. D\"ur, G. Vidal, and J. I. Cirac, 
Phys. Rev. A {\bf 62}, 062314 (2000).
\bibitem{T} The operator $T$ can be obtained based on Eqs.~(10-11) of Ref.~\cite{OS12}.
\bibitem{ES12}  More precisely, a bound has been developed for the convex roof of $\sqrt{\tau}$ in C.~Eltschka and J.~Siewert, Sci. Rep. {\bf 2}, 942 (2012).
This quantity is also an entanglement monotone and, for pure states, it remains invariant under determinant $1$ stochastic local operations and communication (SLOCC). See also 
C.~Eltschka and J.~Siewert, Phys. Rev. Lett. {\bf 108}, 230502 (2012).

\bibitem{SDP} L. Vandenberghe, S. Boyd, SIAM Review {\bf 38}, 49 (1996).
We used YALMIP and SeDuMi. See
J. L\"ofberg, in Proceedings of the CACSD Conference
(Taipei, Taiwan, 2004), p. 284;
 J. F. Sturm, Optimization Methods and Software {\bf 17},
1105 (2002).

\bibitem{SZ02} M. M. Sinolecka, K. Zyczkowski, and M. Kus, Act. Phys. Pol. B {\bf 33}, 2081 (2002).

\bibitem{normalform} F. Verstraete, J. Dehaene, and B. De Moor,
Phys. Rev. A {\bf 68}, 012103 (2003).


%
%
\bibitem{VW02} G.~Vidal and R.~F.~Werner, Phys. Rev. A {\bf 65}, 032314 (2002).
\bibitem{ES13} C.~Eltschka and J.~Siewert, Phys. Rev. Lett.
{\bf 111}, 100503 (2013)


%
%
\bibitem{DF99}  D.~P.~DiVincenzo {\it et al.},
Entanglement of assistance. In {\it Quantum Computing and Quantum Communications} (pp. 247-257) (Springer, Berlin Heidelberg, 1999); arXiv:quant-ph/9803033.

%
%
\bibitem{BW97} B. Schumacher and M.~D. Westmoreland,  Phys. Rev.
A {\bf 56}, 131 (1997).
\bibitem{OV06} T.~J. Osborne and F. Verstraete, Phys. Rev. Lett. {\bf 96}, 220503 (2006).

\bibitem{qfi} C. Helstrom, Quantum Detection and Estimation Theory
(Academic Press, New York, 1976);
A. Holevo, Probabilistic and Statistical Aspects of Quantum Theory (North-Holland, Amsterdam, 1982);
S.L. Braunstein and C.M. Caves, Phys. Rev. Lett. {\bf 72}, 3439 (1994).





\end{thebibliography}
\end{document}